\documentclass[overfull,figures]{epl}
\usepackage{amssymb,amsmath}
\usepackage{epsf,colordvi,graphicx,color,amsbsy}
\usepackage[T1]{fontenc}

\title{``Turbulent'' electrical transport in Copper powders}
\author{E.~Falcon\thanks{E-mail: \email{Eric.Falcon@ens-lyon.fr}, URL: http://perso.ens-lyon.fr/eric.falcon/} \and B.~Castaing \and C.~Laroche}
\institute{Laboratoire de Physique, \'Ecole Normale Sup\'erieure de Lyon\\ UMR 5672 - 46, all\'ee d'Italie, 69 007 Lyon, France}
\shortauthor{E.~Falcon \etal}
\shorttitle{``Turbulent'' conductance in a metallic powder}
\pacs{45.70.-n}{Granular systems} 
\pacs{05.40.-a}{Fluctuation phenomena, random processes, noise, and Brownian motion} 
\pacs{72.80.-r}{Electrical conductivity of specific materials}

\begin{document}
\maketitle

\begin{abstract}
Compressed copper powder has a very large electrical resistance (1M$\Omega$), due to the oxide layer on grains (100$\mu$m). We observe that its voltage-current $U$--$I$ characteristics are nonlinear, and undergo an instability, from an insulating to a conductive state at relatively small applied voltages.  Current through the powder is then noisy, and the noise has interesting self-similar properties, including intermittency and scale invariance. We show that heat dissipation plays an essential role in the physics of the system. One piece of evidence is that the instability threshold always corresponds to the same Joule dissipated power whatever the applied stress.  In addition, we observe long--time correlations which suggest that thermal expansion locally creates or destroys contacts, and is the driving mechanism behind the instability and noise observed in this granular system.
\end{abstract}  

\section{Introduction} For over a century \cite{Branly90}, electrical transport in metallic powders has generated interest \cite{Holm58}. These powders have fascinating properties, such as extreme sensitivity to electromagnetic waves, highly nonlinear $U$--$I$ characteristics, hysteresis, and $1/f$ noise, for which fully satisfactory explanations are still lacking. The experiments presented here were motivated by the work of Kamarinos {\it et al.} \cite{Kamarinos75} on compressed copper powders. These authors observed an insulating-to-conducting transition at rather low pressure-dependent voltages, associated with strong $1/f$ noise. Just above the conduction transition threshold, we observe slow temporal evolution of the powder resistance with a noisy component. Electrical breakdown of the oxide layers on grains has been invoked \cite{Kamarinos75,Vandembroucq97} for the transition, but this explanation is unsatisfactory, for we observe that the noise involves both increasing and decreasing the electrical resistance of the powder.  In this letter, we show that this electrical noise has interesting scale invariant properties. Scale invariance \cite{ScaleInvariance} occurs for various physical signals:  turbulent velocity \cite{Frisch95}, financial stock market data \cite{Sornette03}, earthquake energy release \cite{ScaleInvariance}, or world wide information traffic \cite{Willinger00}. One of the goals of this study is to use the analytic tools developed for turbulent signals in the case of this granular system. We shall focus the comparison on two aspects: The statistics of current increments, $I(t+\tau)-I(t)$, depending on the tested time scale $\tau$, and the correlations between the amplitudes of these increments. This latter will show that the electrical noise in this granular system has a hierarchical organization through the time scales. 

\section{Experimental setup} Two kinds of experiments are performed with commercial copper powder samples of 100 $\mu$m ``spherical particles'' \cite{Goodfellow}: $U$--$I$ characteristics on one hand, and noise and relaxation measurements on the other. In both cases, the powder samples are confined in a polymethylmethacrylate (PMMA) cylinder, of 10 mm inner diameter, capped with two metallic electrodes (stainless steel or brass cylinders).  The container is filled with powder up to a height of 5 mm, roughly corresponding to 500 000 particles. For the $U$--$I$ characteristics, a sensor measures the force applied to the powder through the electrodes. We occasionally embedded two wires inside the powder to check that the resistance is not controlled by the electrode-powder interface.  Generally, before each experimental run, the container is refilled with a new sample of powder. This procedure ensures better reproducibility than simply relaxing the confining pressure and shaking the container. 

\section{$U$--$I$ characteristics} The DC current, $I$, is provided by a Kepco Power Supply (BOP 50-4M). We first apply a static force, $F$, to a new sample. Then we measure the voltage, $U$, across the sample, as a function of increasing values of the current, $I$. A single run typically lasts 10 s.

\begin{figure}[ht!]
\twofigures[width=6.5cm]{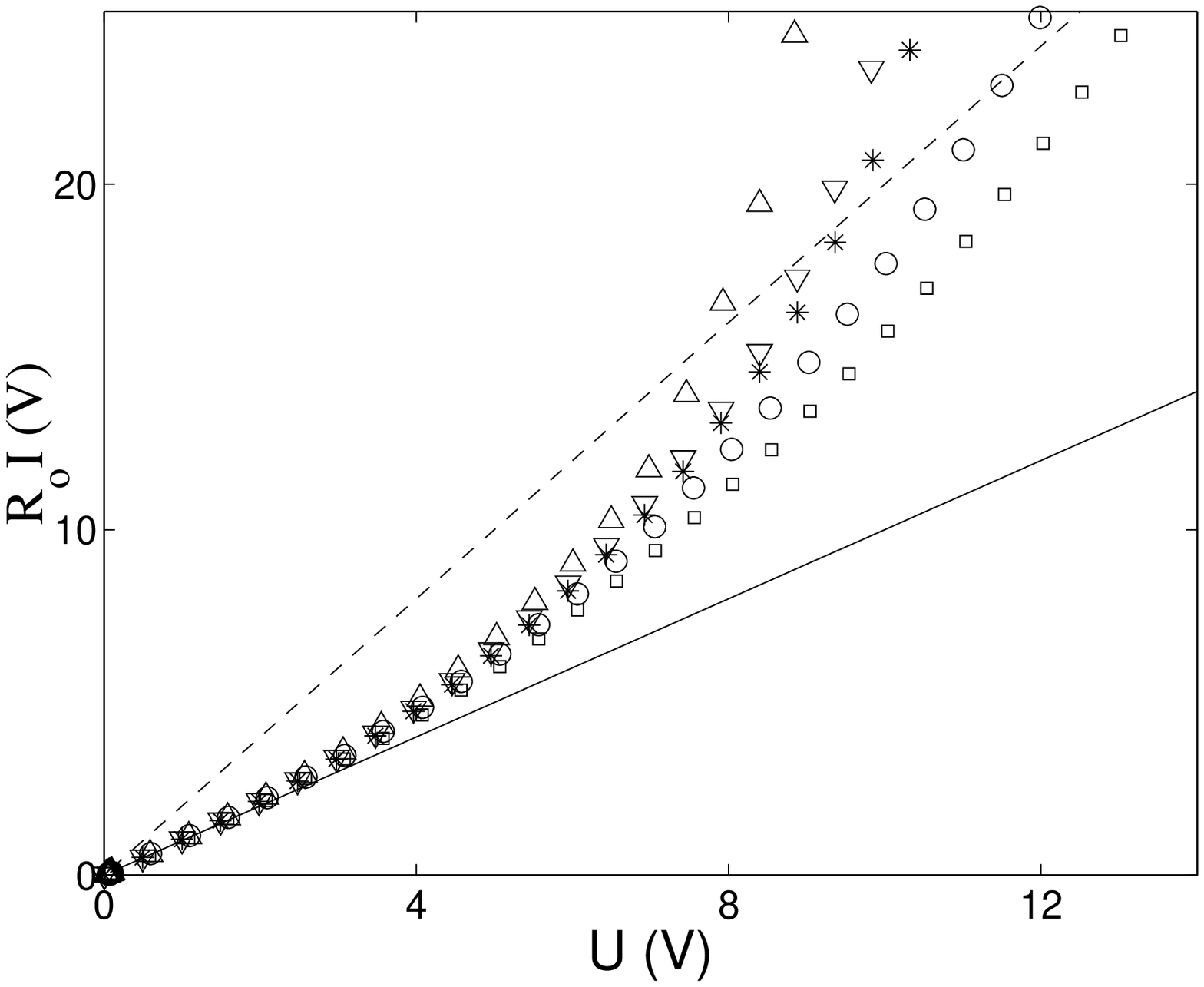}{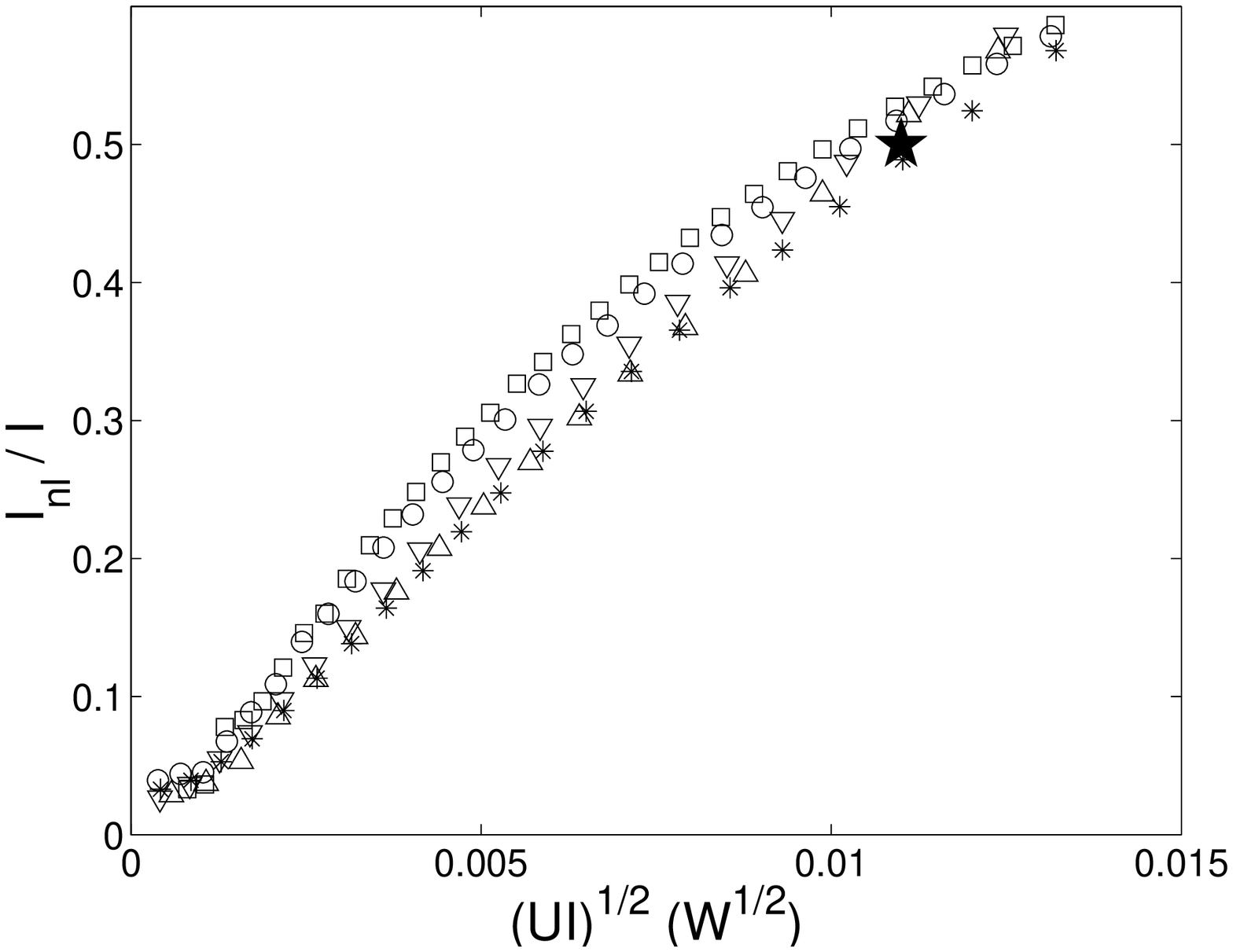}
\caption{Normalized $U$--$I$ characteristics for various applied forces: $F=640$ ($\Box$), 700 ($\circ$), 750 ($\ast$), 800 ($\bigtriangledown$) and 850 ($\bigtriangleup$) N. Slope of unity ($-$). Slope 2 ($--$), {\it i.e.} $I_{nl}/I=1/2$, is close to the instability threshold (see text for details).}
\label{fig.1} 
\caption{The relative nonlinear part of the current, $I_{nl}/I$, versus $(UI)^{1/2}$. The $\bigstar$-mark is the point where $I_{nl}/I=1/2$, ($UI=0.12$ mW ; $UI_{nl}=U^2/R_0=0.06$ mW),  close to the instability threshold (see text for details). Symbols are as in fig.\ \ref{fig.1}.}  
\label{fig.2} 
\end{figure}

Figure \ref{fig.1} displays the $U$--$I$ characteristics for various applied forces ranging from $F=640$ to 850 N. $I$ is normalized by the sample resistance $R_0(F)$ at low $U$, and thus the slope of each characteristic is 1 at the origin. We see that at a higher applied force, the departure from linear behavior occurs at lower voltages (see fig.\ \ref{fig.1}). Note that $R_0(F)$  decreases with $F$, 1M$\Omega$ being a typical value. We define $I_{nl}\equiv I-(U/R_0)$ as the departure from linearity, that is the nonlinear part of $I$. Figure\ \ref{fig.2} then shows that plotting $I_{nl}/I$ as a function of $(UI)^{1/2}$ collapses all data. At first glance, a simple interpretation for this  $(UI)^{1/2}$ dependence can be developed. Indeed, developing $I$ in powers of $U$ such as:
$$
I=(U/R_o) + cU^2 {\rm ,\  \ \ yields\  \ \ } I_{nl}/I \simeq
R_ocU \simeq cR_o^{3/2}(UI)^{1/2} {\rm .}
$$
However, due to the symmetry of the system, $c$ should be zero. Thus, this interpretation does not hold. We prefer to focus on the main observation: The relative nonlinear component of the current, $I_{nl}/I$, seems to depend only on the total dissipated power in the sample, $UI$. An interpretation along these lines will be proposed in what follows.

\section{Transition and relaxation} In a new series of experiments, the PMMA cylinder is filled with 5 mm of powder, then vibrated and embedded in a cylindrical brass press with 5 mm thick walls.  The pressure on the sample is generated by means of the lid press acting as a screw. At constant low voltage ($U\simeq 0.5$ V), the sample resistance, $R_0$,  is monitored during the stepwise pressing of the sample until a maximum pressure, $P_m$, is reached. $R_0$ is found to decrease with increasing $P_m$, and to reach a value ranging from 100 k$\Omega$ to 500 k$\Omega$, at the end of the pressing process (half an hour later). We then let the system relax for one day. After checking that $R_0$ and $P_m$ are constant in time, a fixed voltage, $U$, is suddenly applied to the sample, and the current, $I$, is monitored. If $U$ is smaller than a threshold value $U_c$, the sample is in a weakly conducting state with very slow temporal evolutions of $I$, if any. When $U>U_c$, an instability occurs. $I$ rapidly increases at constant $U$ which can be interpreted as the resistance relaxing down. The larger $U$ is, the faster the resistance relaxes. Similar features have been observed by Kamarinos {\it et al.} \cite{Kamarinos75}. 

When the above experiment is repeated for different values of $P_m$, it shows that both $U_c$ and $R_0$ depend on $P_m$. However, the critical ratio $U_c^2/R_o$ is found to be independent of the applied pressure, with a value close to $0.07$ mW. This value is close to the point where $I_{nl}/I=1/2$ (see $\bigstar$-mark in fig.\ \ref{fig.2}). All these observations suggest that this spontaneous transition from an insulating to conducting state is a thermal instability. To further characterize the phenomenon observed for $U>U_c$, let us make three remarks:

\begin{itemize}
\item Due to the strong sample pressure, vibrations have no effect; even a strong jolt applied to the press has no visible consequence on the signal. 

\item We occasionally followed the relaxation down to a resistance of 500 $\Omega$, which is almost three orders of magnitude lower than the initial one $R_o$; however, this resistance remains much higher than a metallic contact would produce between the grains, no matter how small the contact \cite{Holm58}.   

\item The temporal evolution of the current, at constant $U>U_c$, is not monotonic. Both increasing and decreasing events occur for the current. However, the former dominate and control the global evolution. This is contradictory to what would result from a cascading electrical breakdown for the oxide layers, as supposed by Kamarinos {\it et al.} \cite{Kamarinos75}. 
\end{itemize}

\begin{figure}[ht!]
\twofigures[width=6.5cm]{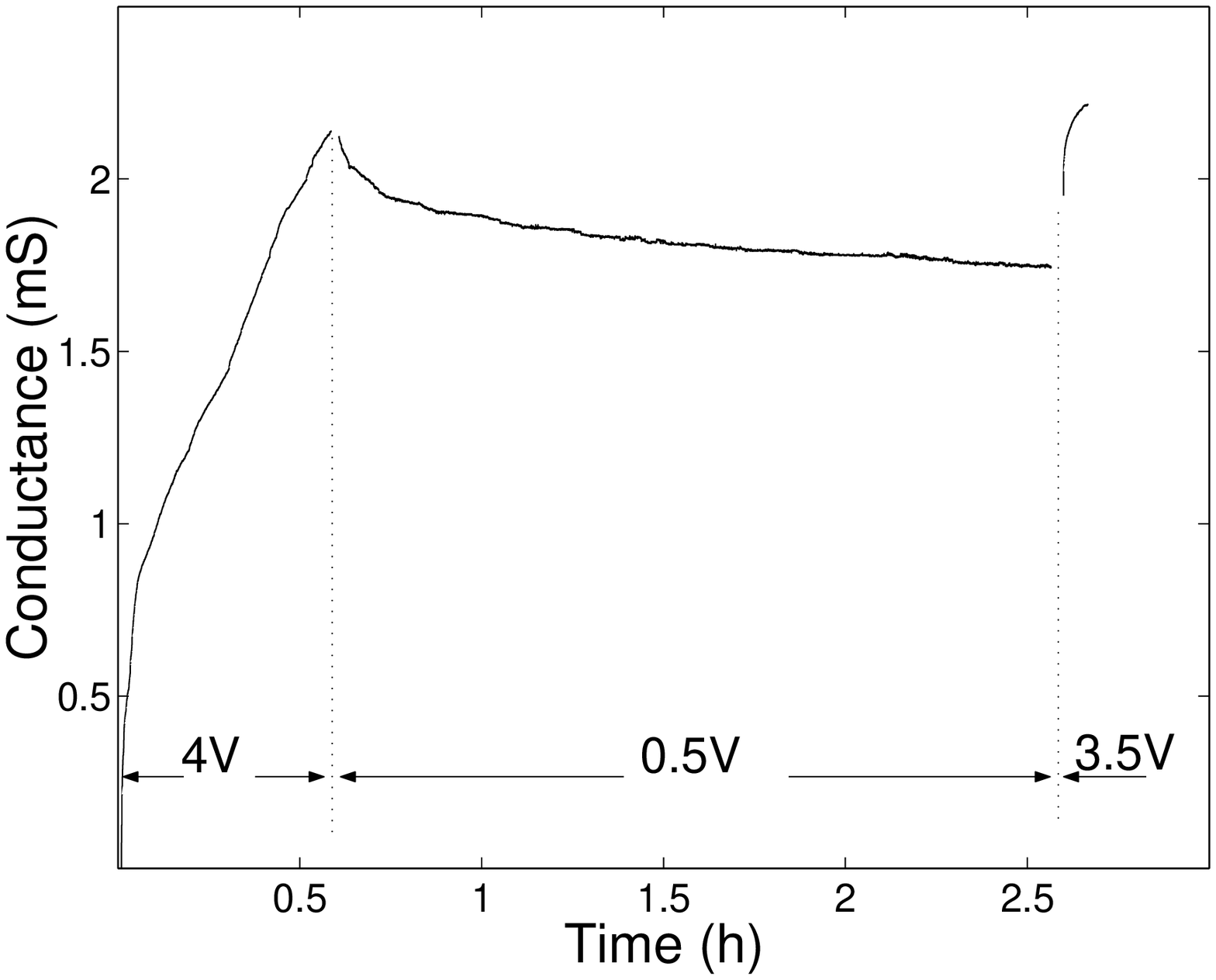}{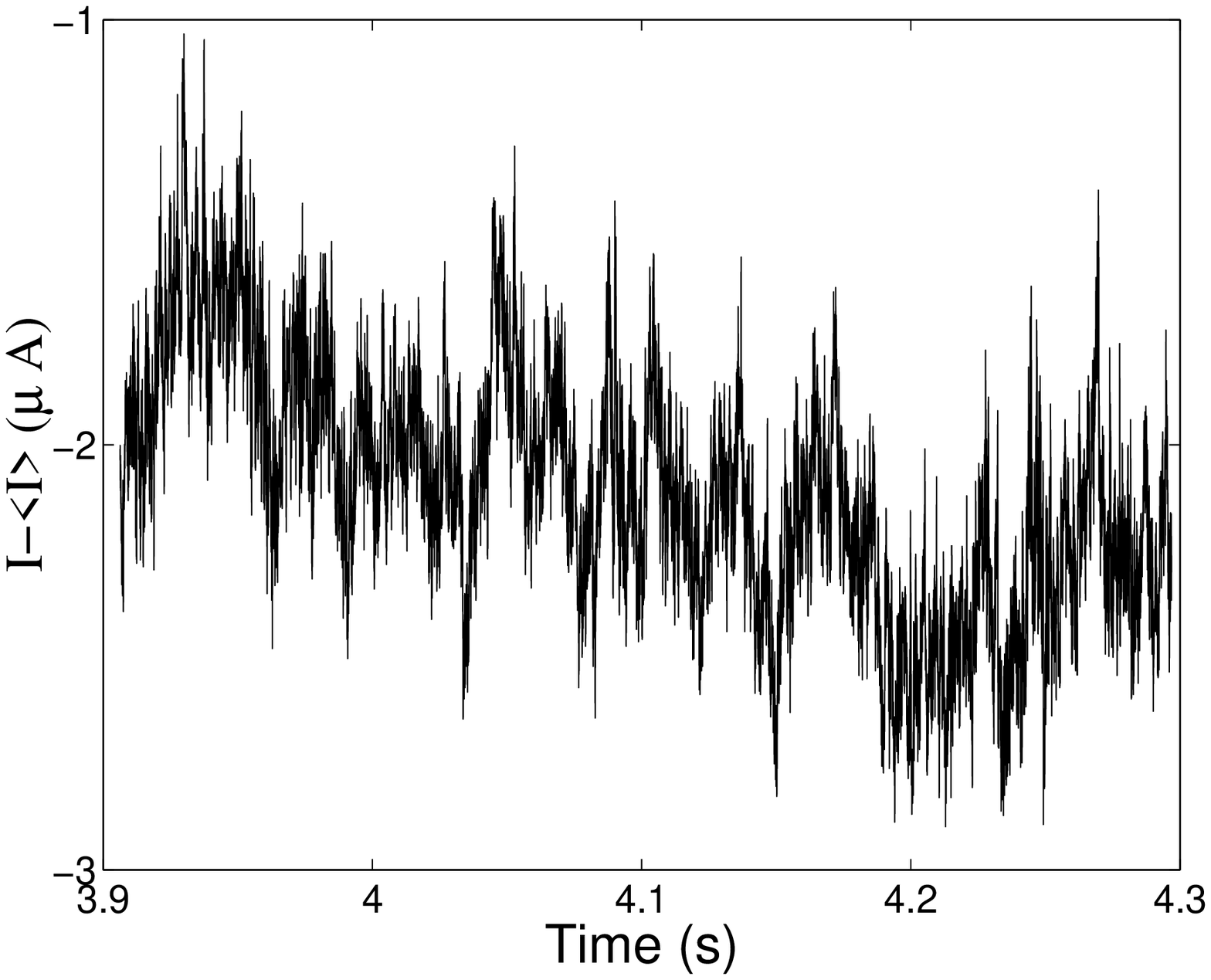}
\caption{Temporal evolution of the sample conductance. The 4V first applied are above the threshold. In the second part, 0.5V are applied, and finally 3.5V.} 
\label{fig.3}
\caption{Quasi-stationary current noise can be obtained with both increasing and decreasing events.} 
\label{fig.4}
\end{figure}

Finally, we observe that the direction of the global evolution of the resistance depends on the applied voltage. As shown in fig.\ \ref{fig.3}, we first apply to the  sample a voltage $U=4$ V ($U^2/R_0=0.1$ mW $>U_c^2/R_0$),  which triggers the relaxation. Half an hour later, we decrease $U$ to 0.5 V, and the conductance goes down. Two hours later, we increase $U$ to 3.5 V and we see that the conductance goes up again. Consequently, with a well choosen applied voltage, we can obtain an approximately constant conductance. Note that these observations are coherent with the last item above. We exploit them in what follows.
 
\section{Noise} Our goal is now to obtain a quasi-stationary current signal in order to apply the usual tools of signal processing ({\it e.g.} spectral analysis), and more sophisticated ones developed in the framework of studies of turbulence time series \cite{Frisch95}. To start, we consider an initial sample resistance of $R_0=0.5$ M$\Omega$. To obtain the relaxation, we apply $U\simeq 7$ V ($U^2/R_0\simeq 0.1$ mW $>U_c^2/R_0$). Five minutes later, the resistance reaches 830 $\Omega$, we then apply $U=0.5$ V, and the resistance goes up, reaching 1 k$\Omega$ one hour later. Applying now $U=2$ V, the resistance slowly decreases to a value of 950 $\Omega$ after one hour. Finally, applying $U=1.6$ V leads to an approximately constant current for hours, with stochastic fluctuations (see fig.\ \ref{fig.4}). At fixed voltage $U=1.6$ V, the current $I$ is recorded through an acquisition system, with a sampling frequency of $f_H=25.6$ kHz (resp. $f_L=128$ Hz), the signal being previously filtered at 10 kHz (resp. 50 Hz) to avoid aliasing. The signal is recorded during 20 s (resp. 65 min) leading to a file of 0.5 Mpt.  This type of data acquisition was repeated 20 times, first at $f_H$, then at $f_L$, to extract averaged quantities, due to the quasi-stationarity feature of the signal. 

\begin{figure}
\twofigures[width=6.5cm]{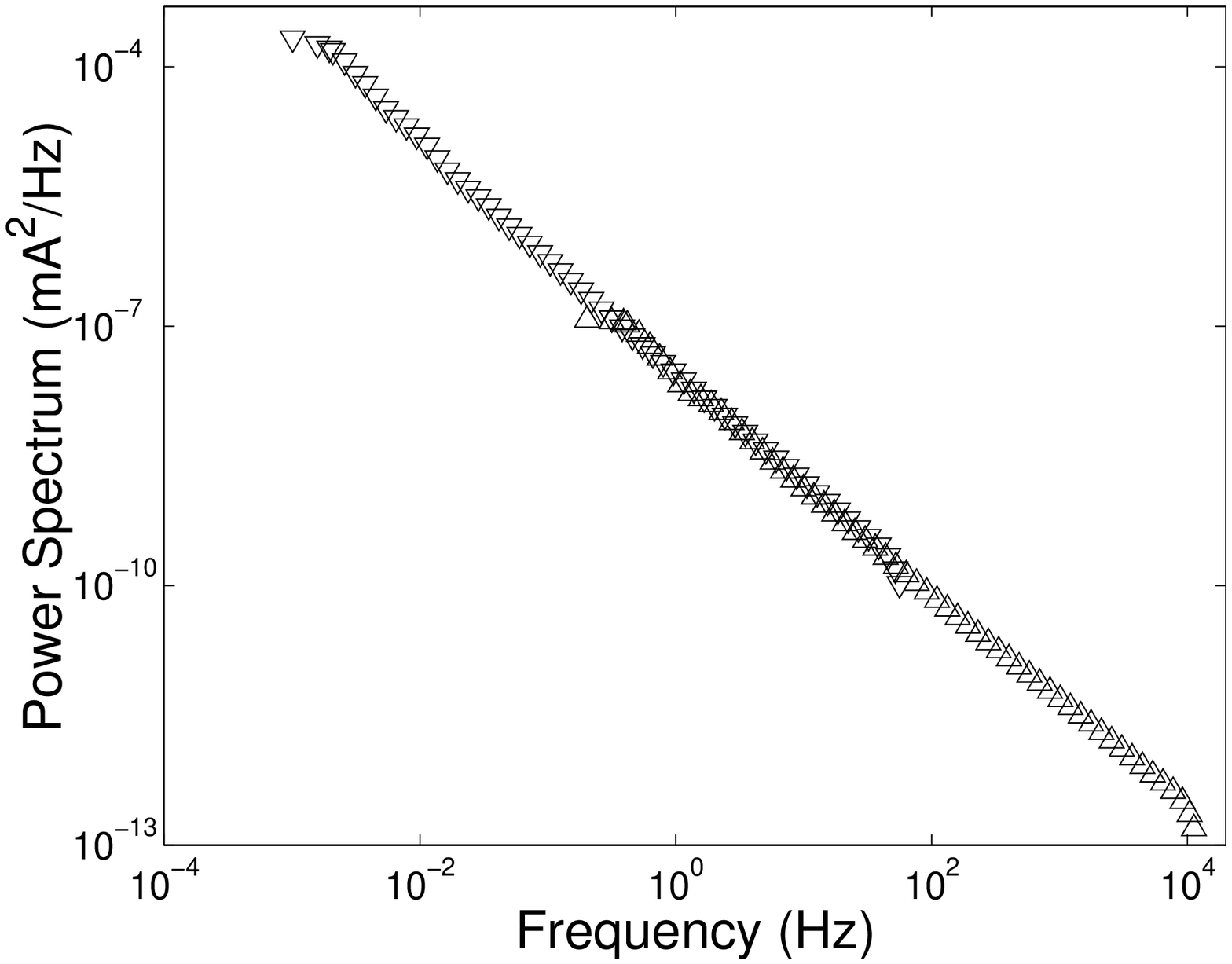}{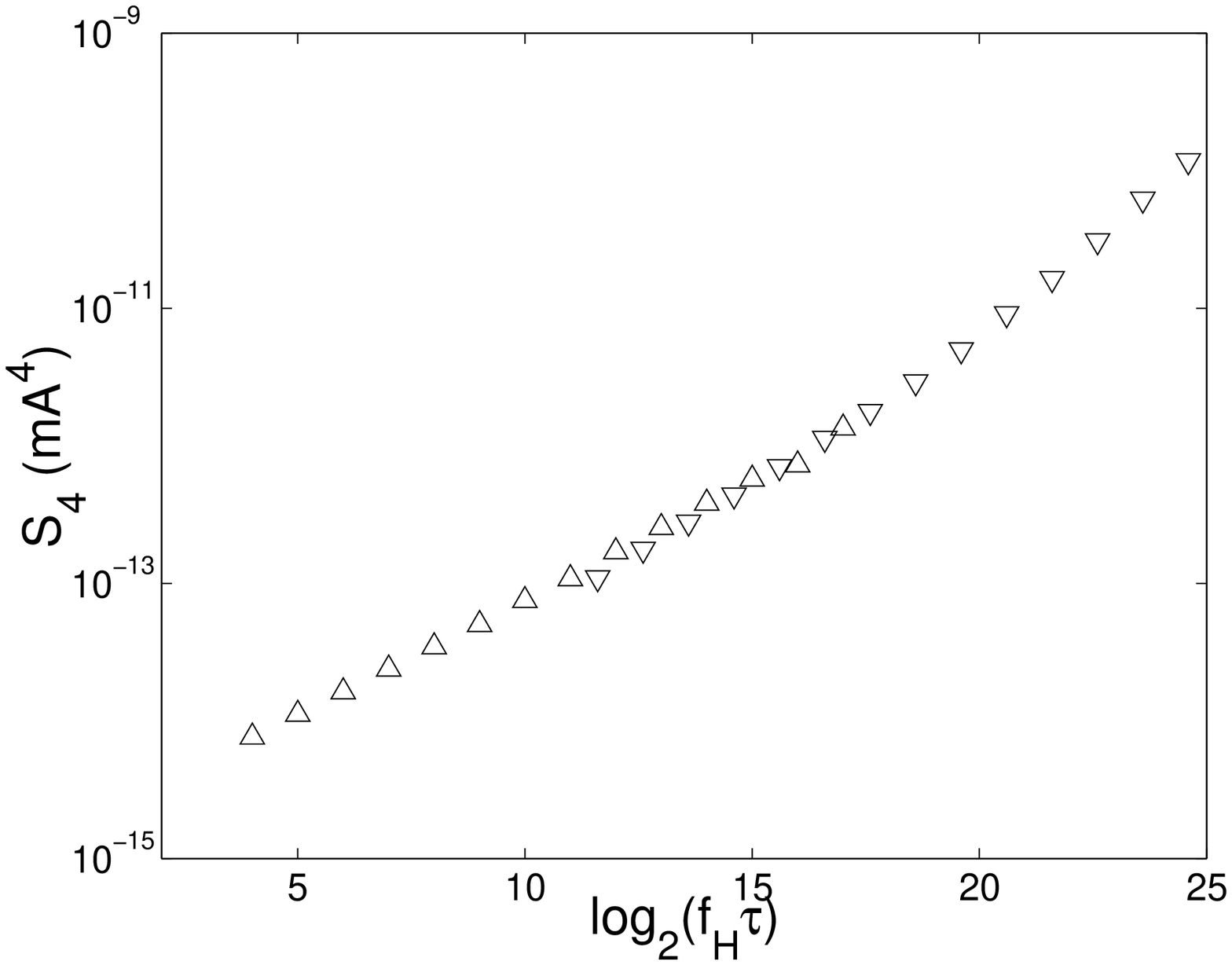}
\caption{Average power spectra of current fluctuations sampled at low ($\triangledown$) and high ($\triangle$) frequencies.}  
\label{fig.5} 
\caption{The fourth moment of current differences, $S_4(\tau)$, versus the nondimensional time scale, $f_H\tau$. Same symbols as fig.\ \ref{fig.5}.}
\label{fig.6}
\end{figure}

Figure\ \ref{fig.5} displays the log-log power spectra of filtered signals of current recorded at $f_L$ and $f_H$. At first sight, this power spectrum of current fluctuations seems to be a power law of the frequency. However, when one examines the spectra carefully, a small curvature appears in fig.\ \ref{fig.5}. Letting $\tau$ designate a time lag or a time scale, we define the {\em i}-th order structure function $S_i(\tau)=\left\langle \left[I(t+\tau)-I(t)\right]^i \right\rangle $, where $\langle \cdot \rangle$ represents an average over time $t$.  We focus on the structure function of order four as a function of $f_H\tau$, as shown in fig.\ \ref{fig.6}. In this log-log plot, high sampling frequency data present a power law dependence with $f_H\tau$. Deviations from this are observed at large times (e.g., $\log_2{f_H\tau} \gtrsim 16$), that is at times $\tau$ greater than a critical time scale $\tau_c \simeq 3$ s. This critical time $\tau_c$ can  be understood as a typical effective diffusive time of a thermal pertubation within our typical size sample, $L\simeq 2.5 - 5$ mm, estimates corresponding to $\tau_c  \simeq 1-10$ s. This order of magnitude agreement supports the hypothesis of a thermally driven phenomenon. With similar reasoning the heat diffusion in a single grain provides the short time limit, $\tau_{inf}  \simeq 0.1$ ms, which is on the order of the inverse of our 10 kHz filtered frequency. For high frequency data, $S_4(\tau) \propto \tau^{\alpha_4}$, with $\alpha_4 \simeq 0.57$. The same data give $S_2(\tau) \propto \tau^{\alpha_2}$, with $\alpha_2 \simeq 0.31$. These power-law behaviors are consistent with scale invariance over more than 3 decades in time. Intermittency, corresponding to $\alpha_4<2\alpha_2$, appears more clearly from the direct examination of probability density functions, and correlations, in the next paragraphs. Here, $\alpha_4$  and $2\alpha_2$ differ only by one standard error. 

\section{Probability density functions} We concentrate now on the high frequency part of the signal, which presents nice scale invariance properties for $\tau_{inf} < \tau < \tau_c$. Figure\ \ref{fig.7} shows, for three different $\tau$, the probability density functions (PDF) of current differences, $\delta I_{\tau}(t)=I(t+\tau)-I(t)$, normalized to their respective root mean square, $\sigma$. The exact shape of these PDF is rather sensitive to the statistics (see PDF tails in fig.\ \ref{fig.7}), and would have led to smoother curves with a greater quantities of data. However, two remarks can be made in the light of what is known for velocity signals in turbulence \cite{Frisch95}:
 
\begin{itemize}
\item The PDF shape changes with the time scale $\tau$ -- this is a direct signature of intermittency, as $S_4$ cannot be proportional to $S_2^2$ and thus $\alpha_4 \neq 2\alpha_2$.

\item Current difference PDF are symmetric -- time reversal, which {\em a priori} should not be invoked here, is the only symmetry able to lead to this behavior. Turbulent velocity differences have a skewed PDF, $ S_3$ being proportional to the dissipated power \cite{Frisch95}. However, even in turbulence, global quantities, like the total dissipated power (equivalent to what we measure here) have rather symmetric time difference PDF \cite{Pinton}. 
\end{itemize}

\section{Multiplicative cascade} A signal with scale invariance is called self-similar since its statistical properties can be described by the same laws at various scales. If the shape of time difference PDF changes across scales with a self-similar law of deformation, this signal is generally described either in terms of a ``multifractal set of singularities'', or a ``multiplicative cascade'' across scales, both approaches being considered equivalent \cite{Chainais01}. In the ``multiplicative cascade'', a given scale (e.g., average gradients on a time interval) conditions smaller scales (e.g., smaller intervals within this one) in a random Markovian way. It induces not only intermittency (evolution of distribution shape across scales), but long-range correlations between short intervals \cite{Chainais01,Delour02}. Therefore, if our current signal can be described by a self-similar multiplicative cascade, then the mean squared deviation of $\ln|\delta I_{\tau}|$ should linearly depend on $\ln\tau$, such that
\begin{equation*} 
\label{c2}
C_2(\tau) = \left\langle\left[\ln|\delta I_{\tau}|-<\ln|\delta I_{\tau}|>\right]^2\right\rangle=-\mu \ln(\tau/T) + c {\rm \ ,} 
\end{equation*}
where $T$ is some large time scale. Also, as shown by \cite{Delour02}, the log-correlation between two short intervals (of size $\theta$) should depend linearly on $\ln\tau$,
\begin{equation*}
\begin{split} 
\label{l2}
L_{\theta}^ 2(\tau) &= \left\langle\left[\ln|\delta I_{\theta}(t+\tau)|\ln|\delta I_{\theta}(t)|-<\ln|\delta I_{\theta}|>^2\right]\right\rangle   \\
			     &= -\mu\ln(\tau/T) + c'  {\rm ,} \\
\end{split}
\end{equation*}
with the same coefficient $\mu$, $c$ and $c'$ being two constants.

\begin{figure}
\twofigures[width=6.5cm]{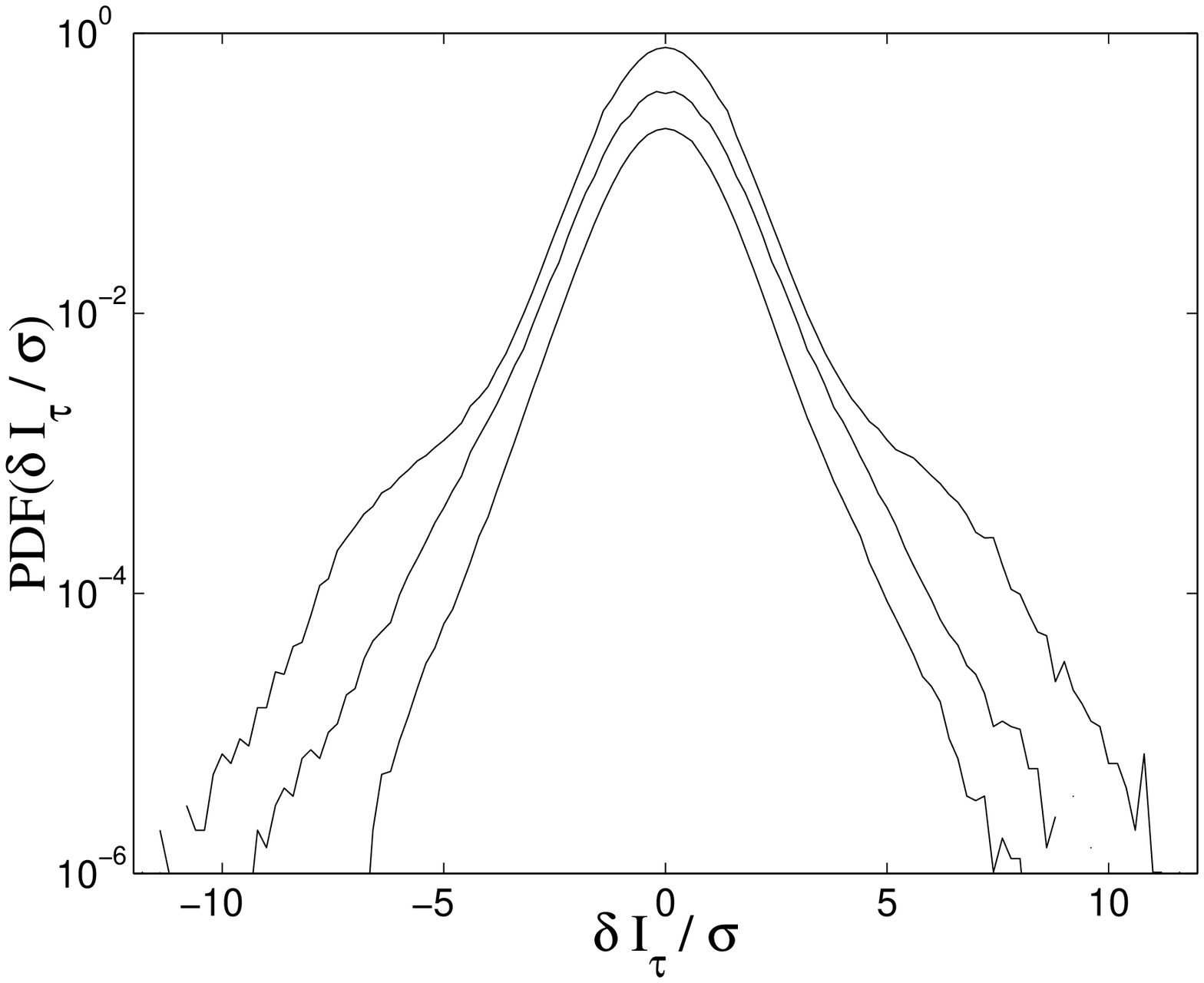}{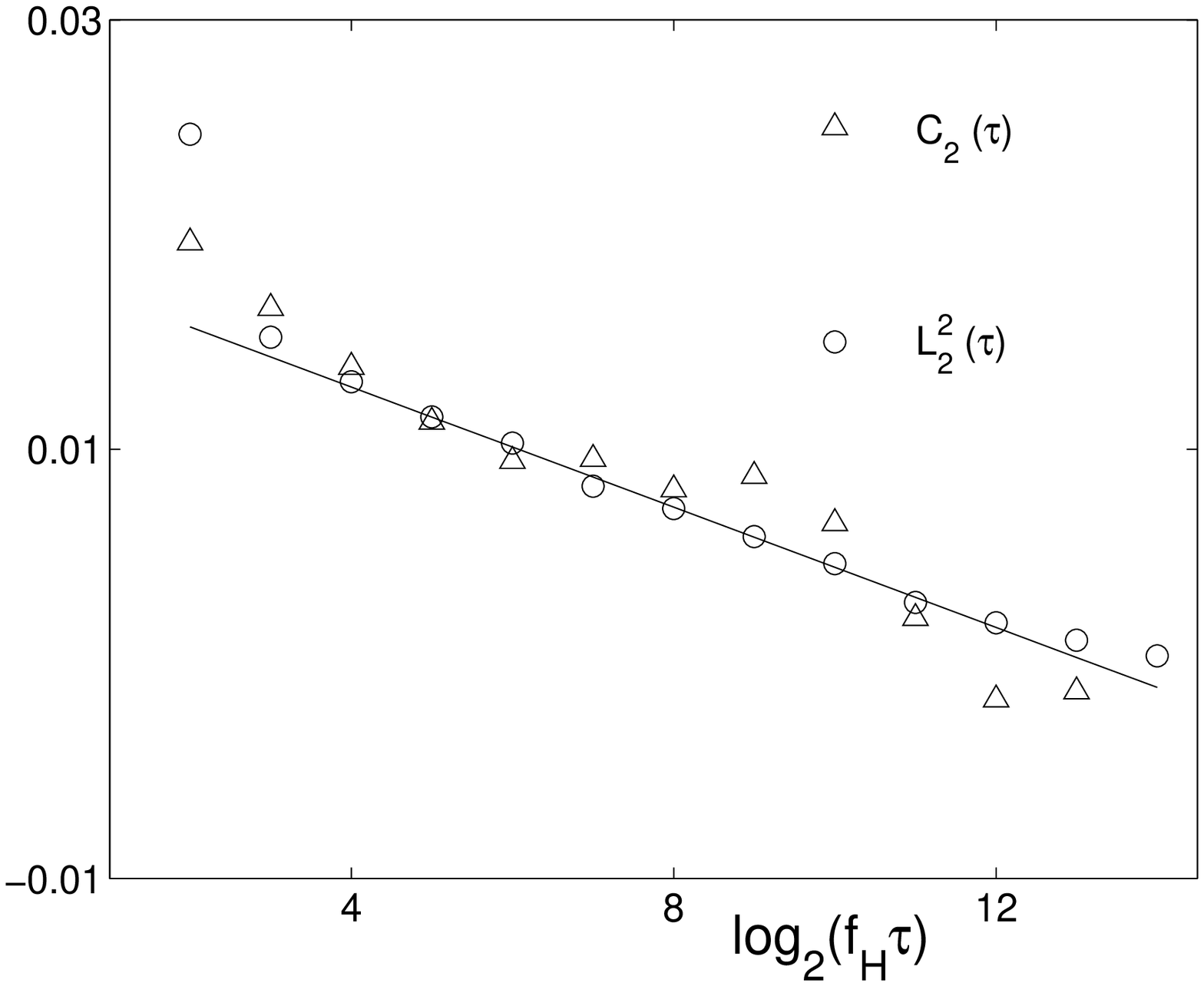}
\caption{Probability density function of the current differences for various time scales: $\tau=$0.15, 2.5 and 40 ms (from top to bottom). Factor of 2 shifts have been applied for clarity.}
\label{fig.7}
\caption{ Comparison  between the variance of the logarithms of current differences ($C_2$) and their correlation ($L^2_{\theta}$ with $f_H\theta=2^2$). Both are close and depend logarithmically on the time separation $\tau$, in agreement with multiplicative cascade models (see text). Solid line of slope $-\mu\ln 2=-1.4\,10^{-3}$ is shown. }  
\label{fig.8}
\end{figure}

Figure \ref{fig.8} shows the two experimental quantities $C_2(\tau)$ and $L_{\theta}^ 2(\tau)$, shifted by appropriate constants. The agreement between them and the linearity in $\ln\tau$ strongly supports a multiplicative cascade description. As in the case of turbulence, large scale events condition those at a smaller scale. Any physical interpretation of the phenomenon discussed in this paper must address these points.

\section{Interpretation} We can now take stock of all of our results and propose a physical picture of what is taking place. As shown above, the driving parameter is the total dissipated power. This suggests either local heating, able to change the electrical properties of contacts (several hundreds of degrees are needed in this case), or thermal expansion, which locally creates or destroys contacts. The power involved is of the order of $10^{-4}$W. When divided by the number of contacts, it remains so small that only electrons can undergo significant heating. Such ``hot'' electron effects have been reported in systems having some analogies with this one \cite{electrons}. However, the influence of the large scales on the small ones, as shown by the observed logarithmic correlations, cannot be taken into account by such a local process. Therefore, thermal expansion seems to be the mechanism driving the instability and the associated noise.

In this spirit, we attempt to show the correspondence between the formal multiplicative cascade and electrical conduction in our powder. Since the contact distribution in a powder is very inhomogeneous, one would also expect an inhomogeneous current distribution. Thus Joule heating should create inhomogeneous increasing stresses in the powder. Very small thermal expansion can result in dramatic changes in the current paths, thus in the distribution of this Joule heating, and so on. Such events can occur at any scale, ranging from the size of the sample and the grain size. The large scale events should influence the small ones, as suggested by our study. A full confirmation of this idea requires longer carefully controlled studies.

\section{Conclusion} The results of this work are two-fold. First, we show that the spontaneous decrease in resistance of a copper powder sample above a voltage threshold is due to a thermal instability, and not to electrical breakdown, as previously had been proposed. This conclusion results from the observation that dissipated power drives the phenomenon, in spite of the probable smallness of the induced temperature inhomogeneities. Second, we propose a procedure yielding an interesting self-similar process in this non-equilibrium system. Our system displays both intermittency and multiplicative cascade-like two point correlations, in ways that are interesting to compare and contrast with the archetypical case of turbulence. 

\acknowledgments  
We wish to thank O.~Michel for his exciting ideas, P.~Metz and D.~Bouraya for electronical and technical support, and G.~Kamarinos, T.~Lopez Rios and L.~K.~J.~Vandamme for discussions. Thanks are also due to V.~Bergeron, L.~Chevillard and M.~Marder for help in improving the manuscript.

\end{document}